%% file: kk258mn1.tex
\documentclass[useAMS,usenatbib]{mn2e}
\usepackage{graphicx, rotating}
\usepackage{aas_macros}  
\usepackage{graphicx}
\usepackage{amssymb}
\usepackage{rotating}
\usepackage{multirow}
\usepackage[breaklinks,colorlinks=true,citecolor=black,linkcolor=black,urlcolor=blue,bookmarks=true]{hyperref}

\newcommand{\kms}{km\,s$^{-1}$}
\newcommand{\ha}{\textrm{H}$\alpha$}
\title[A new transition dwarf galaxy KK258]{KK258, a new transition dwarf 
galaxy neighbouring the Local Group
\thanks{Based on observations made with the NASA/ESA Hubble
Space Telescope, program GO-12546, with data archive at the Space
Telescope Science Institute. STScI is operated by the Associa-
tion of Universities for Research in Astronomy, Inc. under NASA
contract NAS 5-26555.}
\footnotemark[0]\thanks{%
Based on observations made with the Southern African Large
Telescope (SALT).}
}
\author[I. D. Karachentsev et al.]{
I. D. Karachentsev$^{1}$\thanks{E-mail: ikar@sao.ru}, 
L. N. Makarova$^{1}$,
R. B. Tully$^{2}$,
Po-Feng Wu$^{2}$,
A. Y. Kniazev$^{3,4,5}$\\
$^{1}$Special Astrophysical Observatory, Nizhniy Arkhyz, 
Karachai-Cherkessia 369167, Russia\\
$^{2}$Institute for Astronomy, University of Hawaii, 2680 Woodlawn Drive, 
HI 96822, USA\\
$^{3}$South African Astronomical Observatory, PO Box 9, 7935 Observatory, 
Cape Town, South Africa.\\
$^{4}$Southern African Large Telescope Foundation, PO Box 9, 7935 
Observatory, Cape Town, South Africa.\\
$^{5}$Sternberg Astronomical Institute, Lomonosov Moscow State University, 
Moscow, Russia
}
\begin{document}

\date{Accepted XXX. Received XXX; in original form XXX}

\pagerange{\pageref{firstpage}--\pageref{lastpage}} \pubyear{XXX}

\maketitle

\label{firstpage}

\begin{abstract}
Here we present observations with the Advanced Camera for Surveys on 
the Hubble Space Telescope of the nearby, transition-type dwarf 
galaxy KK258 = ESO468-020. We measure a distance of 2.23$\pm$0.05 
Mpc using the Tip of Red Giant Branch method. We also detect \ha{} 
emission from this gas-poor dwarf transition galaxy at the velocity $V_h$ = 92$\pm$5 \kms{}
or $V_{LG}$ = 150 \kms{}. With this distance and velocity, KK258 lies near 
the local Hubble flow locus with a peculiar velocity $\sim$3~\kms{}. We 
discuss the star formation history of KK258 derived from its colour-
magnitude diagram. The specific star formation rate is estimated to be 
log[sSFR] = $-2.64$ and $-2.84$ (Gyr$^{-1}$) from the FUV-flux and
\ha-flux, respectively. KK258 has the absolute magnitude 
$M_B = -10.3$ mag, the average surface brightness of 26.0 mag\,arcsec$^{-2}$ 
and the hydrogen mass ${\rm log}(M_{HI}) < 5.75\,M_\odot$. We compare KK258 with 
29 other dTr- galaxies situated within 5 Mpc from us, and conclude that 
its properties are typical for transition dwarfs. However, KK258 
resides 0.8 Mpc away from its significant neighbour, the Sdm galaxy NGC~55, 
and such a spatial isolation is unusual for the local transition dwarfs.
\end{abstract}


\section{Introduction}

Special sky surveys for dwarf galaxies in the wide vicinities of M~31 
\citep{ibata2007,martin2009,ibata2014} and M 81 \citep{chiboucas2009,chiboucas2013} 
led over a short time to a doubling of the number of dwarf 
companions to these nearest giant spiral galaxies. Significant additions
to the family of dwarf satellites of the Milky Way has been arising over the last
years due mainly to the wide-field sky surveys: Sloan Digital Sky Survey
(SDSS, \citet{abazajian2009}) and Panoramic Survey Telescope and Rapid
Response System 1 (Pan-STARRS1, \citet{tonry2012}) - see a comprehensive
overview by \citet{mconachie2012}.

The sport of hunting for nearby dwarfs is invigorated by the challenge to
resolve the known contradiction between the few observed companions
around local massive galaxies and the great amount of them from theoretical 
expectations \citep{klypin1999}, that exceed by tenfold the observed numbers.
Still, there are observational constraints beyond 1 Mpc
because the limiting apparent magnitude of
the SDSS and Pan-STARRS surveys turns out to be insufficient to resolve
into stars the old population of dwarfs residing outside the Local Group
(LG).

Analysing data on radial velocities and distances of 30 galaxies located
between 1 and 3 Mpc, \citet{karachentsev2009} showed that the Hubble flow
around the LG is rather "cold" with a peculiar velocity dispersion of
$\sigma_V \simeq 25$\,\kms{}. Over the last 5 years, in the spherical
layer within D = 1 -- 3 Mpc only 2 new dwarf galaxies were found:
UGC4879 \citep{kopylov2008, jacobs2011, kirby2012} and Leo P
\citep{Rhode2013, McQuinn2013, Giovanelli2013}. Here, we report on 
the discovery of a third dwarf system,
KK 258, situated at 0.80 Mpc separation from NGC~55, in a low-density 
region between the LG and NGC~253 group in Sculptor.

The interest in such solitary objects is due to an appreciation
that they have not forgotten their "initial conditions"
unlike virialized members of groups. Discovering even one isolated dwarf
with a high peculiar velocity could disprove the coldness of the local
Hubble flow with important cosmological implications.

The low surface brightness dwarf galaxy KK 258 = ESO468-020 was
included in the Neighboring Galaxy Catalog \citep{karachentsev2004} as
having a distance of 3.9 Mpc based on its supposed association with
the giant spiral galaxy NGC 253. However an accurate measurement of the distance
from the luminosities of stars at the tip of the red giant branch 
reveals that KK 258 is well to the foreground, at only half its expected distance. 

\section{ACS HST observations and TRGB distance}

\begin{figure*}
\includegraphics[width=16cm]{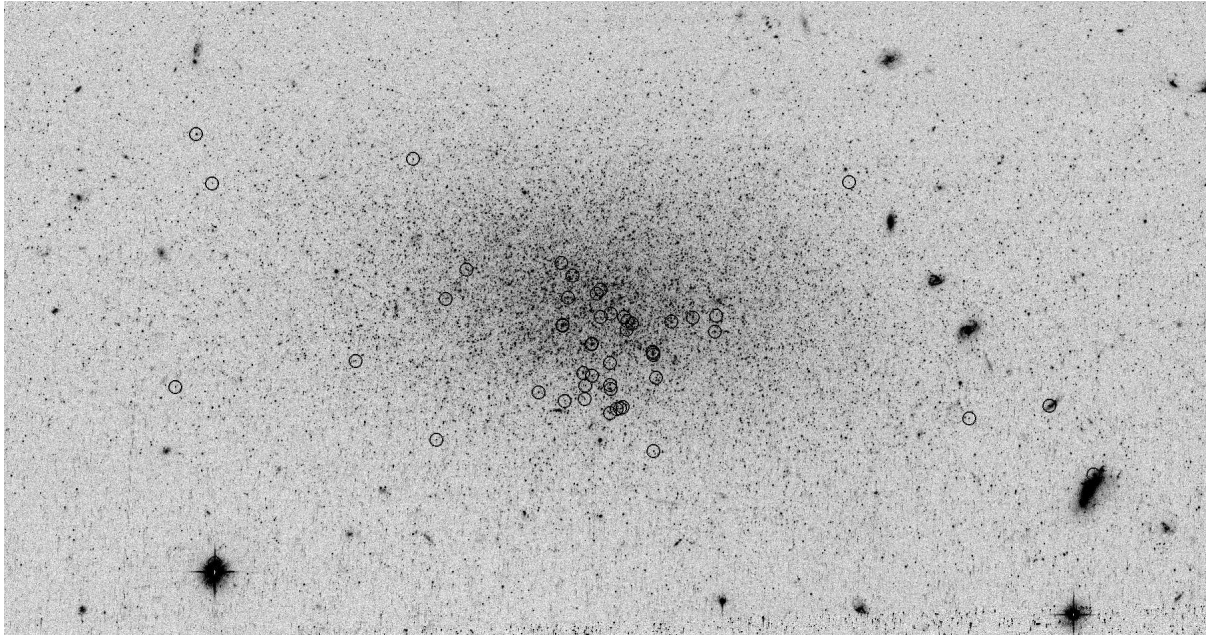}
\caption{\textit{HST}/ACS image of KK258 in \textit{F606W} filter. 
The image size is $3.0\times1.6$ arcmin. Blue stars
with \textit{F606W$-$F814W}$\le$0.2 and \textit{F814W}$\leq$25.5 are shown with open circles.}
\label{fig:ima}
\end{figure*}

\begin{figure*}
\includegraphics[height=10cm]{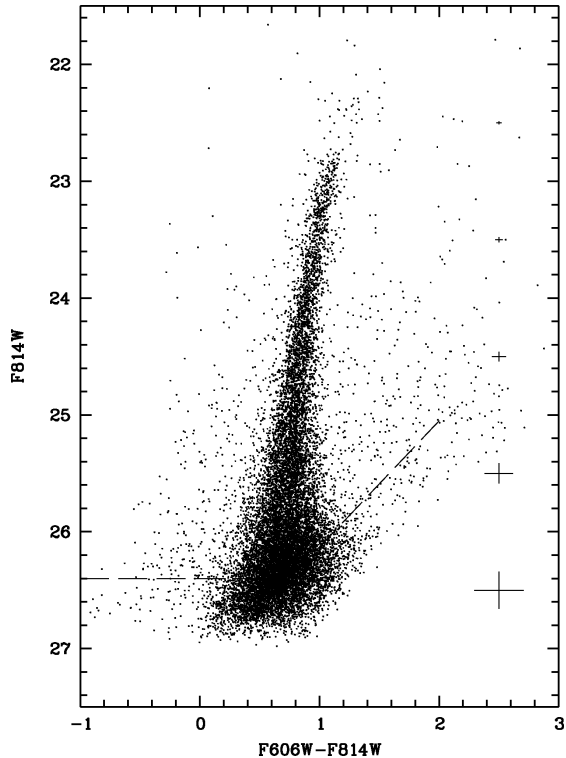}
\caption{Colour-magnitude diagram of KK258. 
The photometric errors are indicated with the bars in the right part of the CMD.
The stars under the dashed line were not used in the SFH measurements.
}
\label{fig:cmd}
\end{figure*}

Observations of KK258 were carried out with the Advanced Camera for Surveys
on the Hubble Space Telescope on August 12, 2012 as a part of SNAP project
12546 (PI: R.B.Tully). Two images were obtained in the F606W and F814W
filters with exposures 900 s in each. The F606W image of the galaxy
is shown in Figure~\ref{fig:ima}.
We use the ACS module of the {\sc DOLPHOT} software package by A.Dolphin
(http://purcell.as.arizona.edu/dolphot) to do photometry of resolved stars
as well as run artificial star tests to characterize the completeness and
uncertainty in the measurements. The data quality images were used to mask
bad pixels. Only stars with photometry of good quality were used in the
analysis. The resulting colour-magnitude diagram (CMD) in F606W -- F814W
versus F814W is plotted in Figure~\ref{fig:cmd}.

 Then, we use a maximum-likelihood method from \citet{makarov2006} to
measure the magnitude of the tip of the red giant branch (TRGB) in KK258
and find F814W(TRGB) = 22.79 $\pm$ 0.04. Following the zero-point calibration
of the absolute magnitude of the TRGB developed by \citet{rizzi2007},
we obtain M(TRGB) = $-4.09$. Using these values and accounting for foreground
reddening of E(B-V) = 0.011 from \citet{sf2011}, we derive a distance
of $(m-M)_0 = 26.85 \pm 0.07$ or D = $2.23 \pm 0.05$ Mpc.

\section{Metallicity and star formation history}

We determine the quantitative star formation and metal enrichment history 
of KK258 from its CMD ({\sc StarProbe} program). Our program 
determines an approximation of the observed distribution of stars in the CMD, using 
a positive linear combination of synthetic diagrams formed by simple stellar 
populations (SSP). A full description of the details of our approach and 
the algorithm are given in the papers of \citet{mm4, makarova2010}.

\begin{figure*}
\includegraphics[width=12cm]{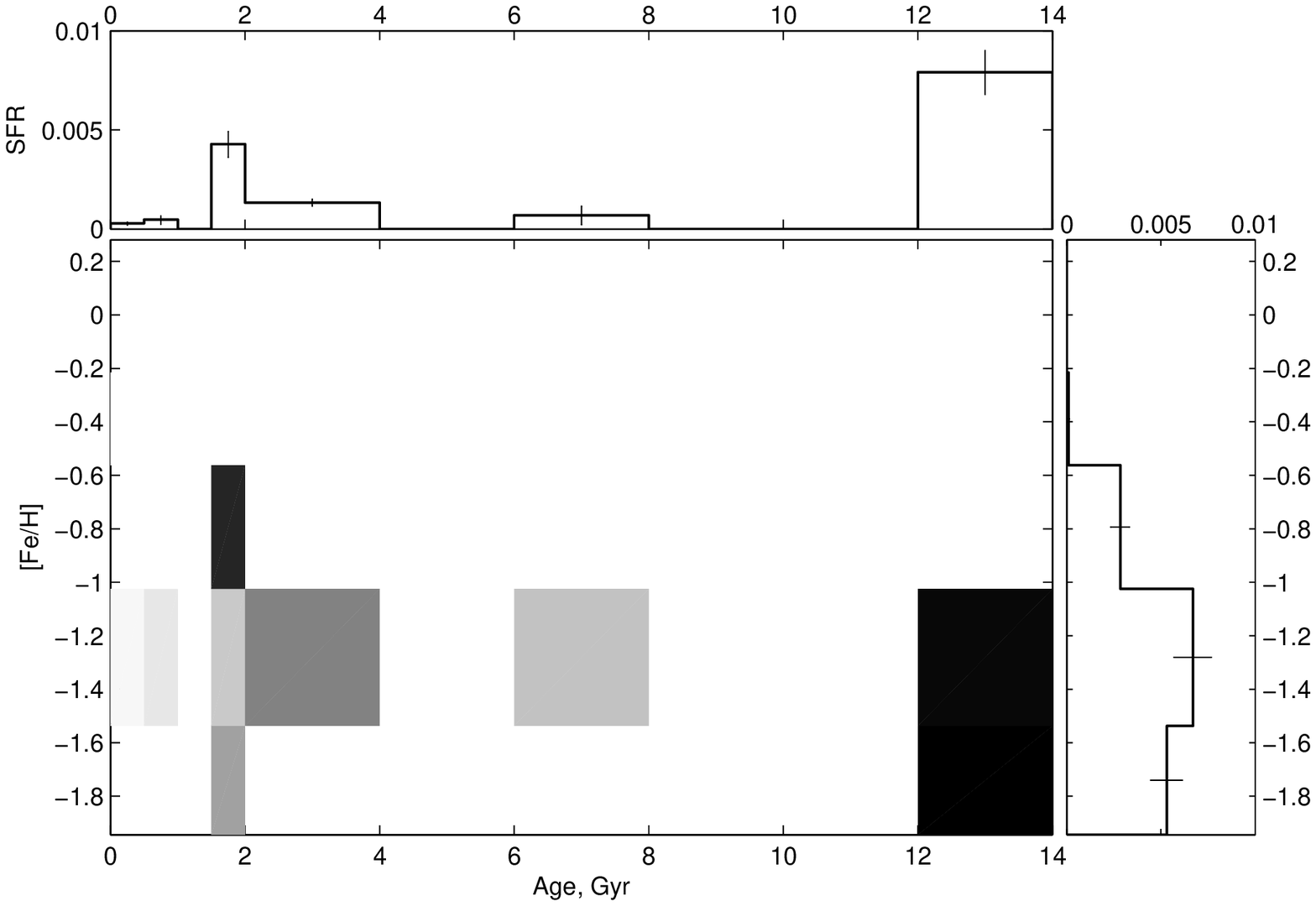}
\caption{The star formation history of KK258.
Top panel shows the star formation rate (SFR) ($M_\odot$/yr) against the
age of the stellar populations.
The bottom panel represents the metallicity of stellar content as a function of
age. The right panel is SFR vs. metallicity.
The resulting errors in SFR are indicated with the vertical bars.
}
\label{fig:sfh}
\end{figure*}

The observed data were binned into Hess diagrams, giving the
number of stars in cells of the CMDs.
Also synthetic Hess diagrams were constructed from theoretical stellar 
isochrones and initial mass function (IMF). 
We used the Padova2000 set of theoretical isochrones \citep{girardi00}, and a 
\citet{salpeter55} IMF. The synthetic diagrams were altered by the same
incompleteness and crowding effects, and photometric systematics
as those determined for the observations using artificial star 
experiments. We used the distance modulus determined in the present paper.
We have taken into account the presence of unresolved binary stars
(binary fraction). Following \citet{barmina}, the binary fraction was taken 
to be 30 per\,cents. The mass function of individual stars and the main component 
of a binary system is supposed to be the same. 
The mass distribution for the second component was taken to be flat 
in the range 0.7 to 1.0 of the main component mass.
The best fitting combination of synthetic CMDs is a maximum-likelihood solution 
taking into account the Poisson noise of star counts in the cells of the Hess diagram.
The result of our calculations of the star formation history of
KK258 is shown in Figure~\ref{fig:sfh}.

The main starburst in KK258 apparently occurred 12 -- 14 
billion years ago. The average rate of star formation in this period was 
$7.9_{-3.2}^{+4.8}\times$10$^{-3}\,M_{\odot}/yr$. The average metallicity 
of the stars formed is quite low, [Fe/H]$\simeq-1.6$. The bulk 
of the stellar mass of the galaxy was formed during this initial starburst. 
We estimate the mass fraction formed during this initial event to be 70 per cent. 
The total stellar mass formed during the galaxy life amounts to $2.2\times$10$^7\,M_{\odot}$. 
Faint traces of star formation are indicated in the time intervals of [6 -- 8] 
and [2 -- 4] Gyr ago. A marked enhancement of star formation likely occurred in 
a short period 1.5 -- 2 billion years ago. The metallicity of these stars is 
slightly higher: [Fe/H] $\simeq-1.0$.

A very weak but reliable sign of star formation is seen in the last billion years.
The average star formation rate in this recent period is 
$3.8_{-1.0}^{+3.0}\times$10$^{-4}\,M_{\odot}/yr$. A very small 
number of stars continue to be formed at the current time.

\section{\ha{} emission and radial velocity}

In the course of a survey of nearby galaxies in the \ha{} line
with the 6-meter telescope of the Special Astrophysical Observatory, Russian
Academy of Sciences, \citet{kk2013} imaged KK258 in the \ha{} line
and in the neighbouring continuum. 
This southern object culminated at the zenith distance of 74 
degrees (!). After the standard data processing, a faint emissive patch of ~5 arcsec size
was detected in the central region of KK 258.  The
integral flux from the patch amounts to $F(H\alpha) = (4.3\pm0.9) 10^{-15}$ erg\,cm$^{-2}$\,s$^{-1}$,
that at the distance of 2.23 Mpc corresponds to the star formation rate
${\rm log}[SFR] = {\rm log}F(H\alpha)+2{\rm log}D+8.98=-4.68$ in units of [M$_\odot$/yr] \citep{kennicutt98}.
This emission knot in the centre of KK 258 is also seen
in ultraviolet images obtained with the space GALEX telescope \citep{gil07}.
The apparent FUV-magnitude of this knot, m$_{FUV}=20.0$\,mag, 
after a small correction for the Galactic extinction, yields the star 
formation rate of ${\rm log}[SFR]=-4.48$ M$_\odot$/yr in good agreement with
the estimate from  \ha{} emission. 
 
Spectroscopic observations of the \ha{} knot in the KK258 were conducted with the SALT 
telescope \citep{Buck06,donoghue06}
on October 23, 2013 with the Robert Stobie Spectrograph \citep[RSS;][]{Burgh03,Kobul03}.
To detect the narrow faint \ha{} line the high resolution long-slit spectroscopy mode
of the RSS was used with a 2.0 arcsec slit width.
The grating GR2300 was used to cover the spectral range 
6100$-$6900 \AA\ with a reciprocal dispersion of $\sim0.27$ \AA\ pixel$^{-1}$
and FWHM spectral resolution of 1.39$\pm$0.05 \AA.
The seeing during the observation was $\approx$2.5 arcsec.
The RSS pixel scale is 0.127 arcsec and the effective field of view is
8 arcmin in diameter. We utilized a binning factor of 4 in the spatial direction,
in order to give a final sampling of 0.51 arcsec pixel$^{-1}$.
The total exposure time was 1950s, which was broken up into 3
subexposures, 650s each, to allow for removal of cosmic rays.
A spectrum of Ne comparison arcs was
obtained to calibrate the wavelength scale.
Primary reduction of the data was done with the
SALT science pipeline \citep{Cr2010}.
The long-slit reduction was done later, in
the way described in \citet{Kn08}.
We would like to note that SALT is a telescope with a variable pupil,
and its illuminated beam changes continuously during the observations.
This makes the absolute flux/magnitude calibration impossible,
even using spectrophotometric standard stars.

\begin{figure*}
\includegraphics[width=7cm,angle=-90]{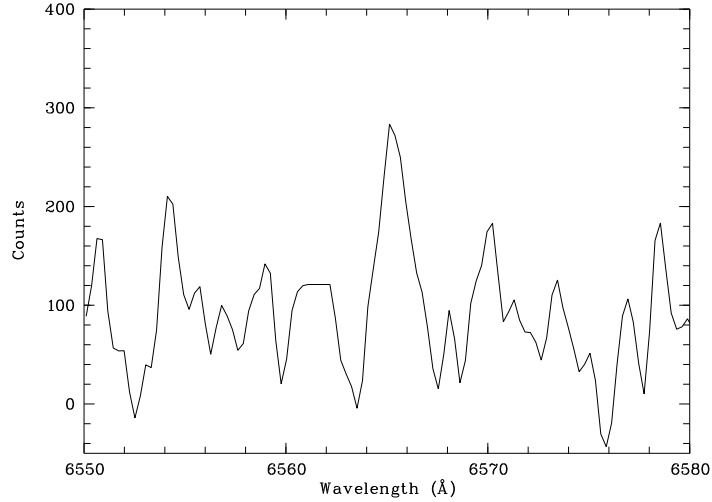}
\caption{One-dimensional scan of the spectrum of KK258.}
\label{fig:spectr}
\end{figure*}

As seen from the data, there is a faint emission line with equivalent
width of $EW \simeq 3.3 $\AA\, at the wavelength 6565 \AA. The one-
dimensional scan of the spectrum in the range of 6550--6580 \AA\,
is presented in Figure~\ref{fig:spectr}. Identification of this emission detail with
the \ha{} line yields the heliocentric velocity 
$V_h = 92\pm5$ \kms{} or the Local Group centroid velocity 
$V_{LG} = 150$ \kms{}.

\section{KK258 and the Local Hubble flow}
\input{table1.tex}

 The data on galaxies that outline the Hubble flow around the LG are
presented in Table 1. The nearby galaxies are selected using two
conditions: a) distance of the galaxy from the LG centroid is
within 3.5 Mpc, b) tidal index of the galaxy $\theta_1$ is less
than +0.2, that avoid selecting objects in groups with appreciable virial
velocities. The data are taken from UNGC catalogue \citep{karachentsev2013b}
which contains references to the data sources given in Tables 4 - 7.
For several objects: SexB, SexA, NGC3741 and DDO210 their UNGC-distances 
were replaced by values from the Extragalactic Distance Database
\citep{Tully2009,Jacobs2009}. The Table columns contain:
(1) galaxy name, (2),(3) equatorial coordinates at the epoch (J2000.0), (4)
distance from the MW and its error, (5) distance from the LG barycenter
assuming its location on the line between the MW and M31 at a distance of
0.43 Mpc away from the MW \citep{karachentsev2009}, (6) heliocentric
radial velocity in \kms{}, (7) velocity in the LG rest frame with
the apex parameters adopted in the NASA Extragalactic Database (NED,
http://ned.ipac.caltech.edu), (8) the tidal index 

$\Theta_1=max[{\rm log}(M^*_n/D^3_n)]-10.96$, n=1,2,...,N,
where $M^*_n$ is a stellar mass of n- rank neighbour, and $D_n$ is its
separation from the considered galaxy \citep{karachentsev2013b}.
Apart from 35 galaxies listed in Table 1, there are three more galaxies
with low velocities $V_{LG}$: 199 \kms{} (UGC6757),
196 \kms{} (LVJ1213+2957), and 145 \kms{} (UGC8245), however,
their distances remain unmeasured.

\begin{figure*}
\includegraphics[width=12cm,angle=-90]{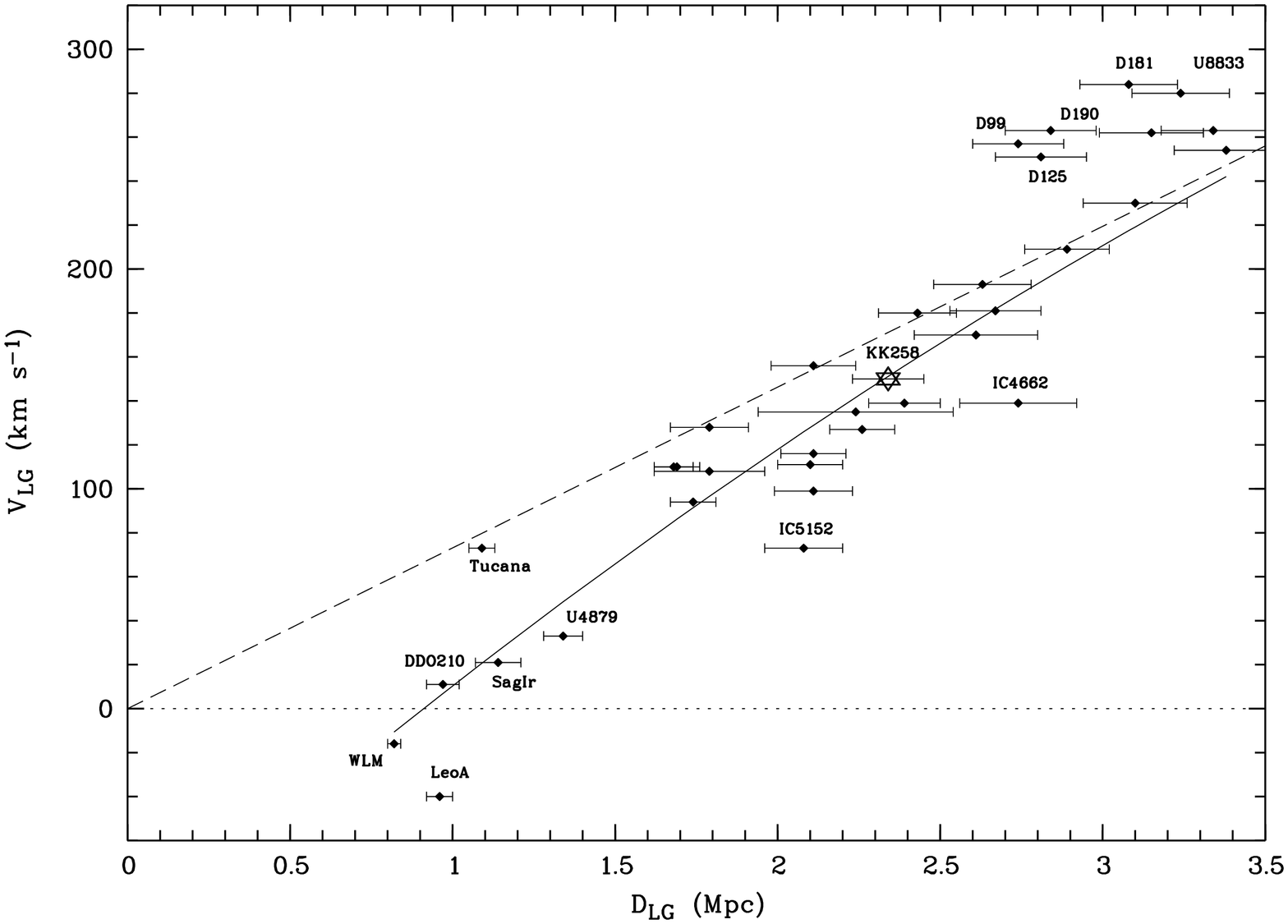}
\caption{Relation between radial velocities and distances of the 35 nearest
isolated galaxies.}
\label{fig:distvel}
\end{figure*}

The relation between the radial velocities and distances of the 35 nearest
isolated galaxies is shown in Figure~\ref{fig:distvel} by solid circles with horizontal
bars indicating the distance errors. The dwarf system KK 258 as
a new member of the local Hubble flow is distinguished by a star.
The dashed line corresponds to the unperturbed Hubble flow with a
parameter H$_0$ = 73 \kms{}\,Mpc$^{-1}$. The solid line represents the
regression of radial velocity on distance for a canonic Lemetre-Tolman model with
the parameter R$_0$ = 0.9 Mpc, that indicates the radius of the sphere
of zero velocity around the LG. As seen, KK 258 lies close to the
regression line. It has a peculiar velocity only
$V_{pec} \sim 3$ \kms{}, comparable with the distance measurement
error of $\sigma_D H_0\simeq8$ \kms{} and the radial velocity error
of 5 \kms{}. 

Note that several galaxies on the local Hubble diagram: DDO 99, DDO 125,
DDO 181, UGC 8833, and DDO 190, members of the dwarf associations 14+7 and 14+8 (Tully et al. 2006),
reside in the Canes Venatici constellation
and exhibit an effect of bulk motion toward the nearby CVn-I cloud 
with a velocity of ~55 \kms{}. For the remaining galaxies the 
peculiar velocity scatter is only 28 \kms{}, and about 10 \kms{}
of this quantity is caused by the distance errors.

\section{KK258 vs. other nearby transition dwarfs}

According to a common point of view, a "transition" type dwarf is a
galaxy dominated by the old population but manifesting signs of recent
star formation. This definition implies the knowledge of detailed star 
formation history derived from a reasonably deep CMD. However, the deep CMDs
are available mostly for the nearest dwarfs.
Given the set of attributes $-$ low hydrogen content, low rate of 
recent star formation
and low surface brightness $-$ the dwarf system KK~258 belongs to the 
transition type, intermediate between irregular (dIr) dwarfs and
spheroidals (dSph). There is a widespread view that the dwarf
transition (=dTr) systems are forming from dIrs via their gas depletion,
whereupon star formation fades. It is usually 
considered that a dying dwarf of the dTr- type transforms with
time into a spheroidal dwarf if it is not fed by gas
from the intergalactic medium. Individual properties of some nearby dTr
dwarfs have been discussed by \citet{hidalgo2013} and \citet{yang2014}.
For obvious reasons, classification of intermediate morphology dwarfs 
faces substantial difficulties. One can cite many cases where dTr dwarfs
were attributed  to the dIr or dSph type.
An example of questionable classification is the case of nearby dIr
galaxy UGC 4879 attributed to dTr type by \citep{kirby2012}.
Another examples of difficulty classifying dTr are the nearby
dwarf galaxies Cetus and Tucana. Based on the luminosity- metallicity 
relation for the Local Group dwarfs, \citet{yang2014} attributed Cetus 
and Tucana to the transition type, while \citet{monelli2010, weisz2011, 
hidalgo2013} described them as typical dSphs because of their negligible 
current star formation rate. On the other hand, \citet{Karachentsev2011}
found a case when a dwarf galaxy DDO44, usually classified as dSph, contains 
a small emission knot with a dozen blue stars producing a star formation 
rate of dex(--5.1) $M_\odot$/yr.

\input{table2.tex}

The UNGC catalogue contains 30 galaxies classified as dTr within a
distance of 5 Mpc. Classification of such objects at greater
distances becomes uncertain. A summary of the nearest dTr
dwarfs is presented in Table 2. The columns represent the following
data from UNGC: (1) galaxy name, (2) equatorial coordinates (J2000.0), 
(3) distance in Mpc, (4) method used to derive the distance: TRGB -
via the tip of the red giant branch, SBF - via surface brightness
fluctuations, mem - via the group membership, txt - via texture of
galaxy image, (5) absolute B-magnitude corrected for the Galactic
extinction according to \citet{schlegel1998}, (6) average surface
brightness within the Holmberg isophote, (7) logarithm of stellar
mass derived from the K-band luminosity, (8) logarithm of 
hydrogen-to-stellar mass ratio, (9) star
formation rate in $M_\odot$/yr, (10) specific star formation rate
in $M_\odot$/yr normalized by the stellar mass, (11) the tidal 
index determined by the nearest significant neighbour, (12) the
group to which the dwarf galaxy belongs. The last line in Table 2
presents median parameters for the dTr sample. The typical transition 
dwarf has the absolute magnitude $M_B=-10.8$ and a stellar mass 
${\rm log}M_*=6.8$. Apparently, galaxies with such a shallow potential 
well easily loose their gaseous component during the first immersion 
into the dense regions of the massive neighbour. Judging from the
median quantity $\theta_1$=0.9, the majority of transition dwarfs 
are in the close neighbourhood of their main disturbers. The typical
hydrogen-to-stellar mass ratio for these galaxies does not exceed
10 per cent, and the observed low star formation rate can only reproduce
$\sim1/30$ part of the stellar mass of dTrs over the cosmic time 
T$_0$=13.7 Gyr. Among the 30 objects of the sample, two dwarf systems:
Phoenix and Tucana are peripheral satellites of the MW, three dTrs:
Cetus, LGS-3 and And XXVIII belong to the M 31 suite, and eight
transition dwarfs are members of the M 81 group.

\begin{figure*}
\includegraphics[width=6cm,angle=-90]{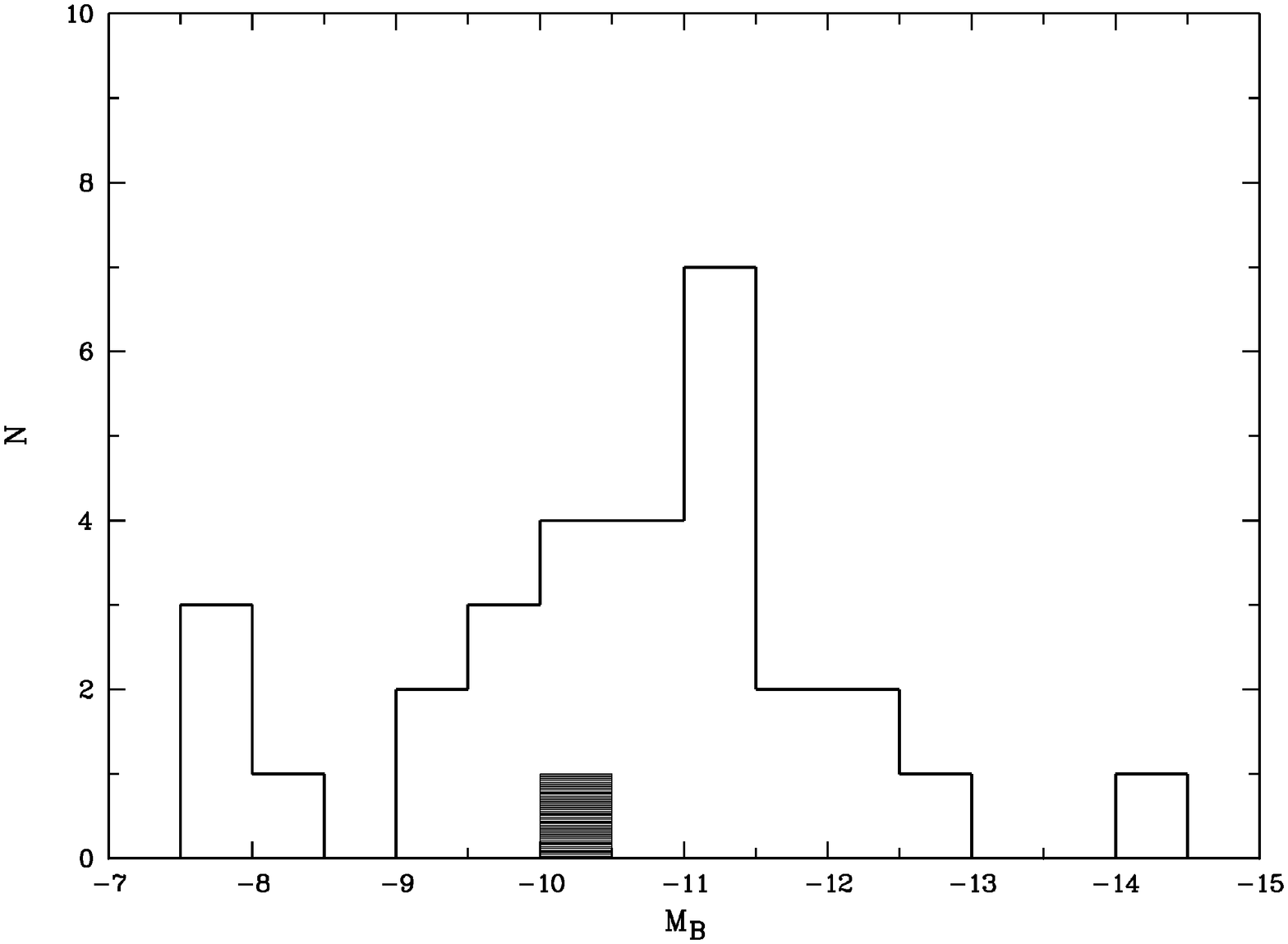}
\includegraphics[width=6cm,angle=-90]{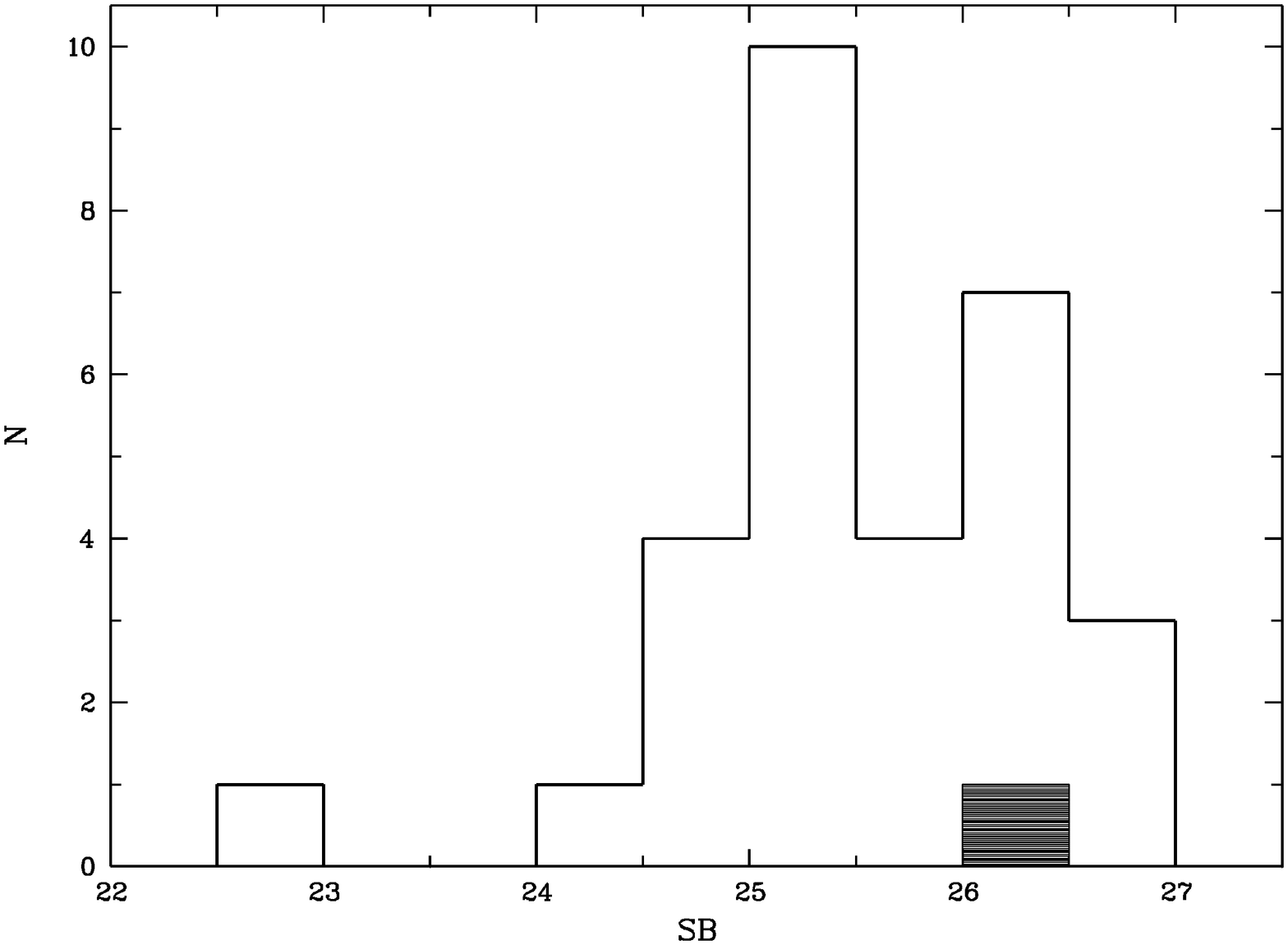}
\includegraphics[width=6cm,angle=-90]{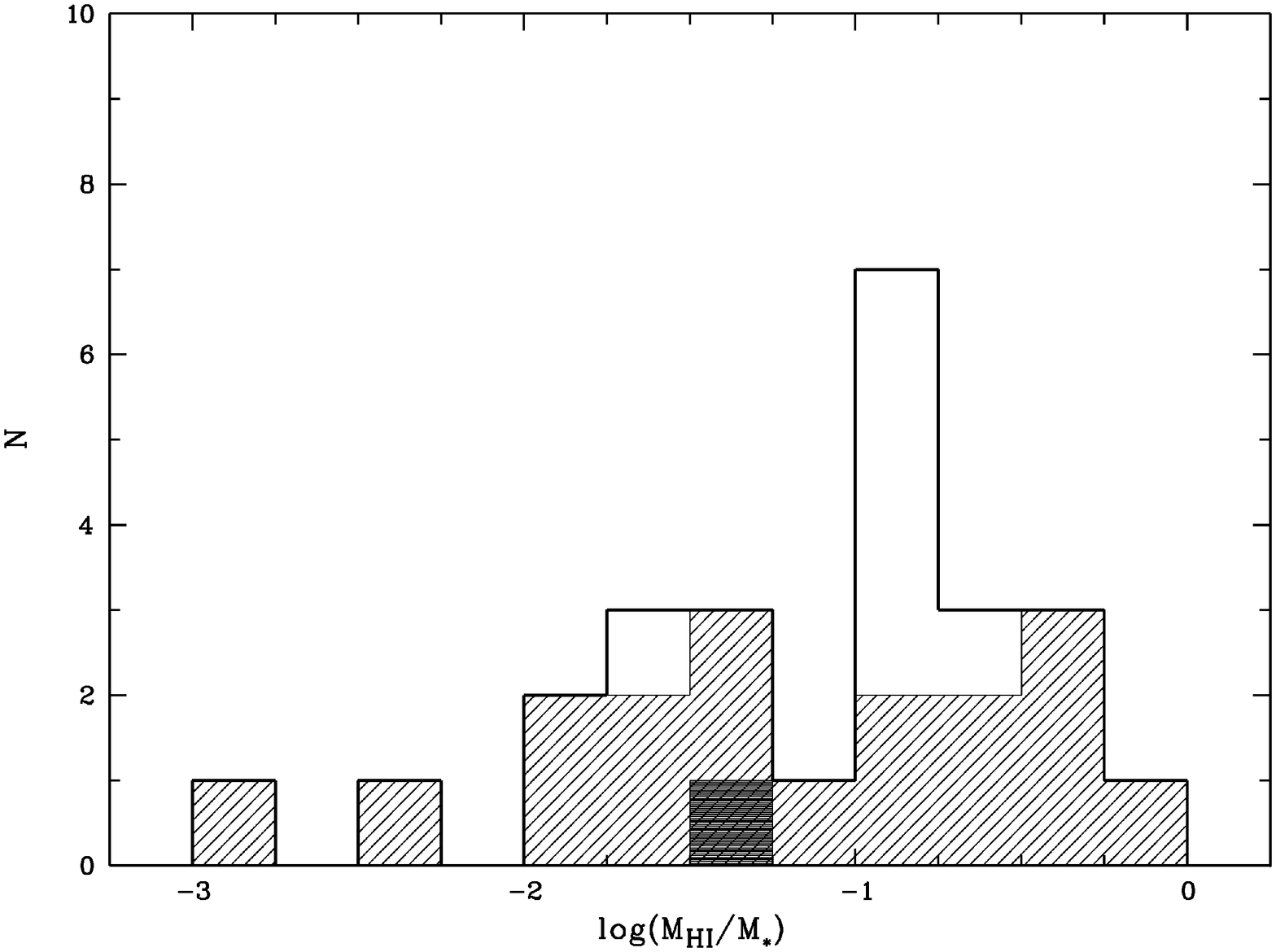}
\includegraphics[width=6cm,angle=-90]{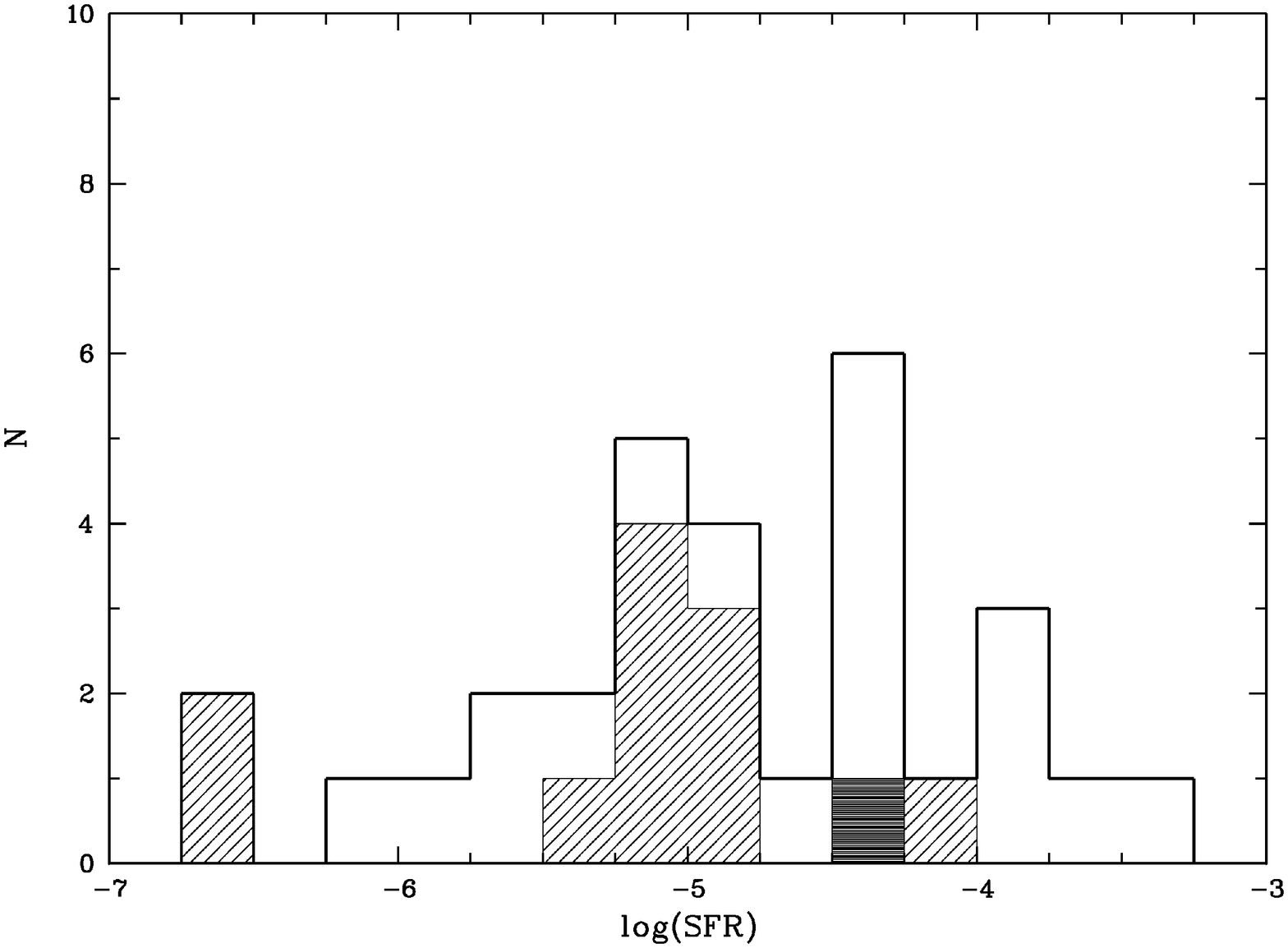}
\includegraphics[width=6cm,angle=-90]{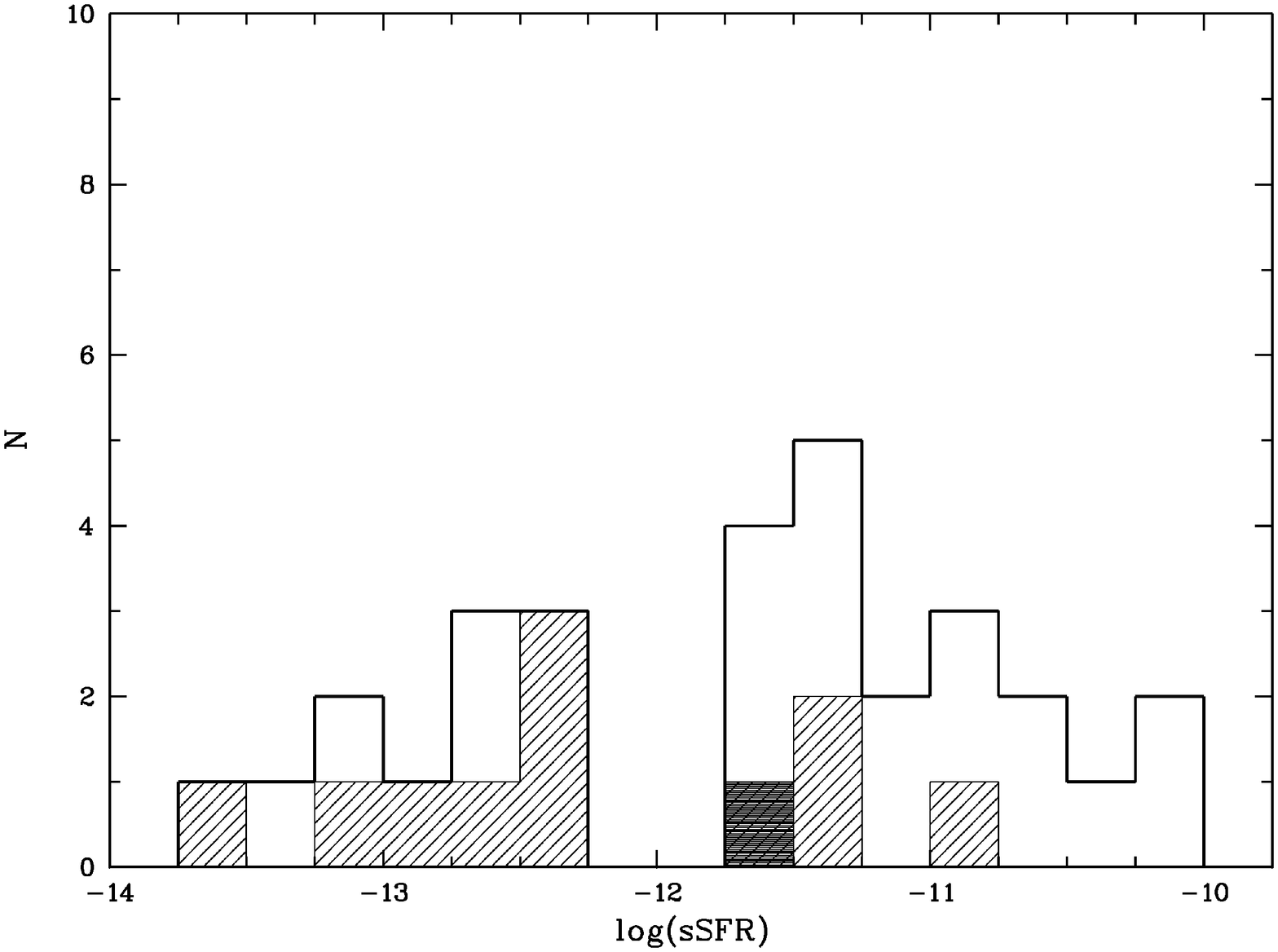}
\includegraphics[width=6cm,angle=-90]{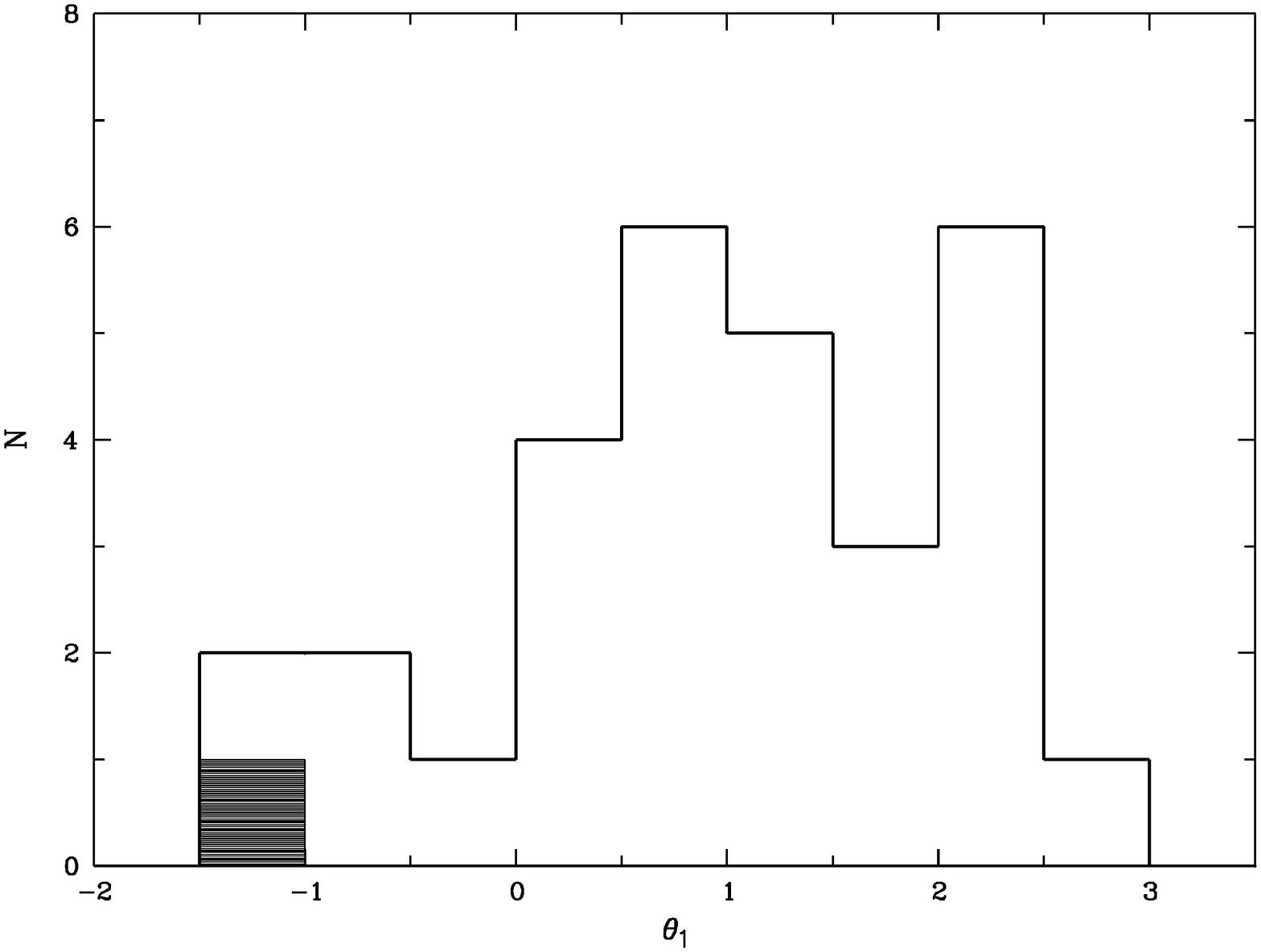}
\caption{A mosaic of histograms showing the distribution
of 30 dTr galaxies according to different parameters.
KK 258 is marked in black. Upper limits of M(HI) and SFR are shaded}
\label{fig:mosaic}
\end{figure*}

Figure~\ref{fig:mosaic} presents a mosaic of histograms which reproduce 
distributions
of 30 dTr galaxies according to different parameters. The galaxy KK 258
is marked in black. The upper limits for M(HI) and SFR are shaded.
KK 258 looks like a
typical representative of the sample according to the majority of parameters:
luminosity, surface brightness, relative hydrogen abundance and star
formation rate. The only difference of KK 258 from other 
dTr dwarfs appears in its tidal index, $\Theta_1 = -1.1$. In the UNGC, the nominally nearest 
significant neighbour to KK 258, i.e. its Main Disturber, is NGC 253 
but, as discussed in the next section, good distances reveal that these galaxies are not associated. 

\section {KK 258 environment}

\begin{figure*}
\includegraphics[width=11cm,angle=-90]{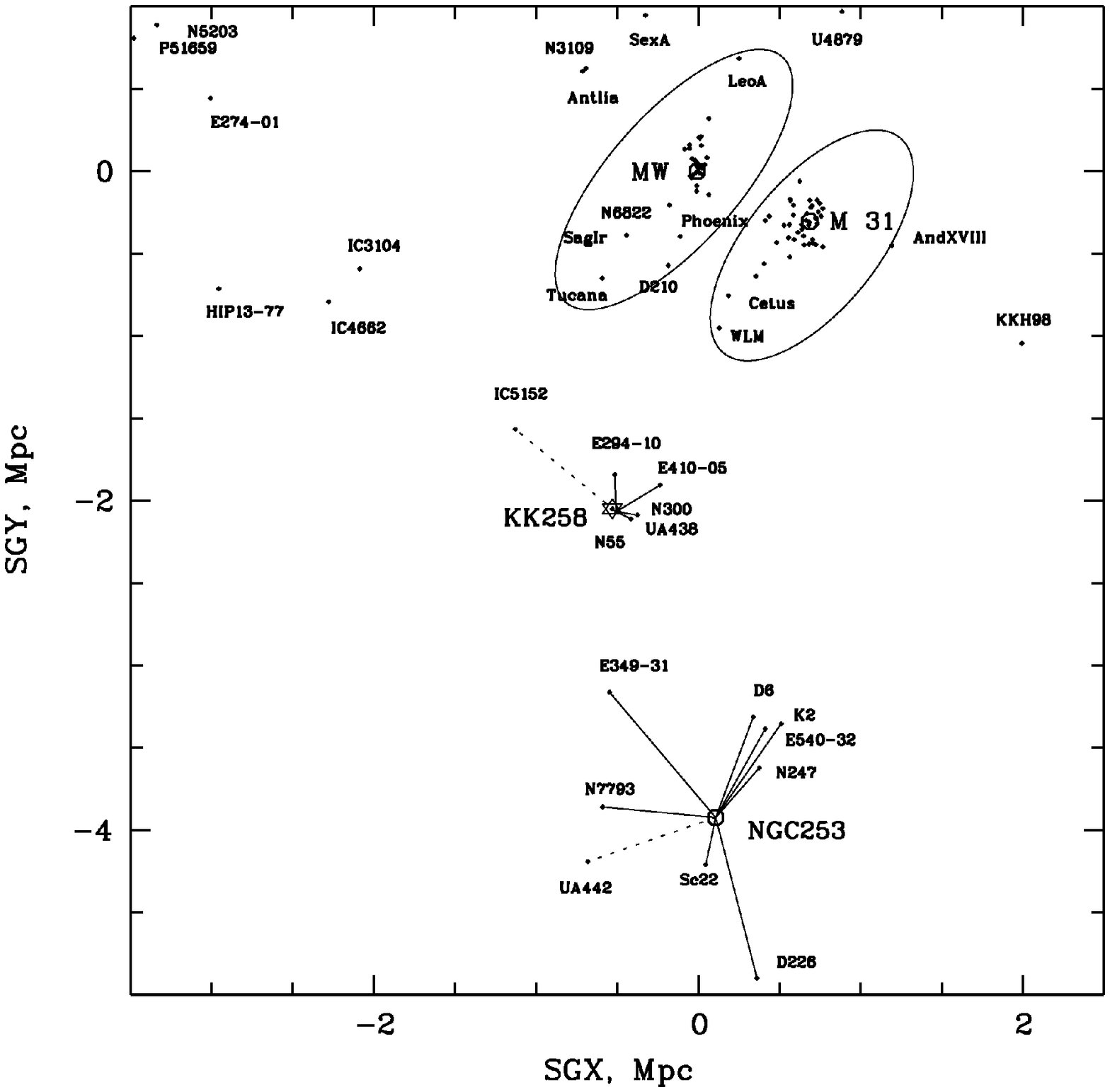}
\includegraphics[width=11cm,angle=-90]{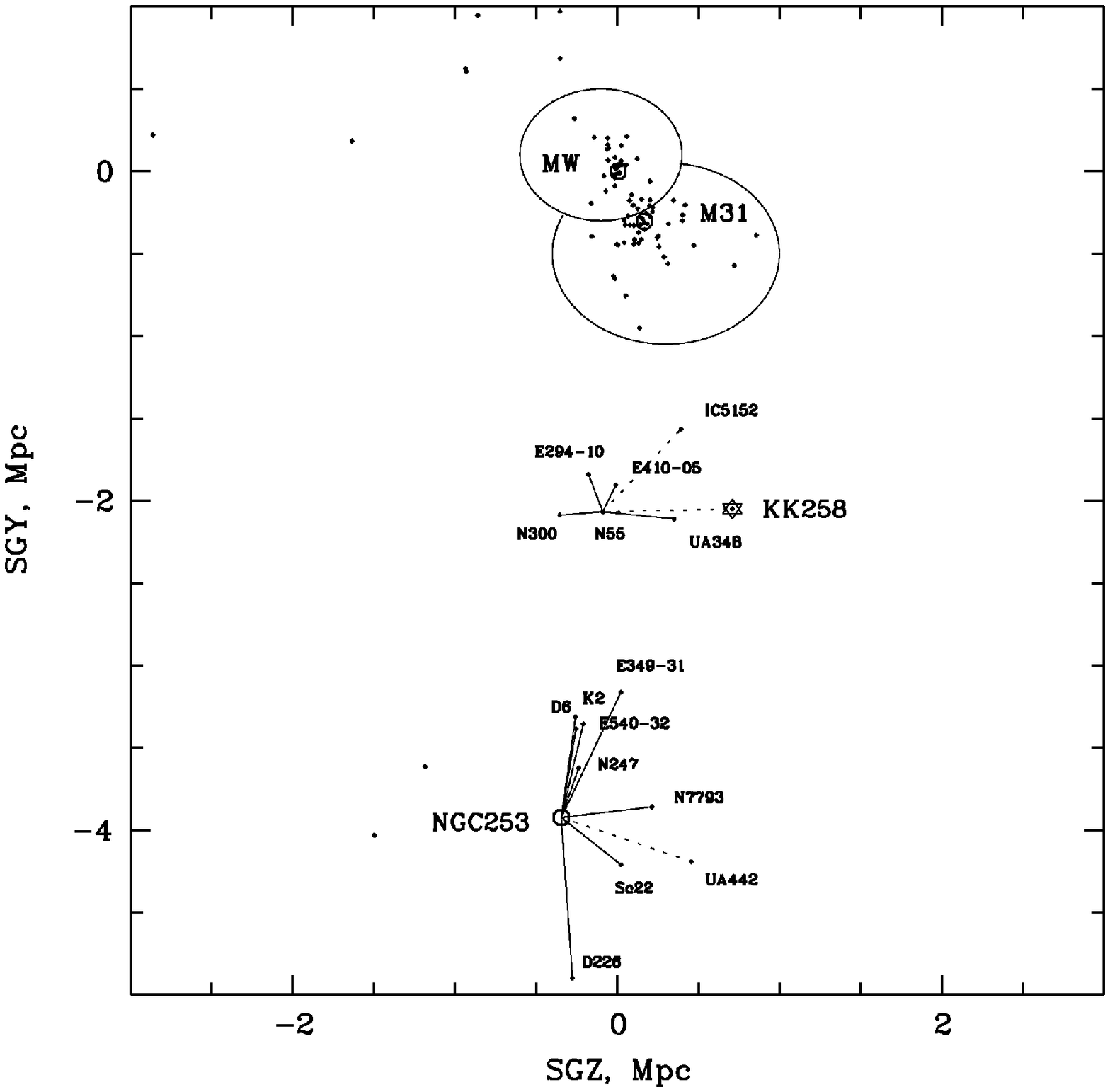}
\caption{The distribution of known galaxies within a distance of 3 Mpc 
around the NGC 55 group in Supergalactic coordinates.}
\label{fig:SGXYZ}
\end{figure*}

Figure~\ref{fig:SGXYZ} demonstrates the distribution of known galaxies within a
distance of 3 Mpc around the NGC 55 group in Supergalactic coordinates.
The galaxies are shown by solid circles with their names labelled. 
Three giant galaxies: the Milky Way, M 31 and NGC 253
having absolute magnitudes M$_B < -$20.0 mag are distinguished by large
open circles. The dwarf object KK 258 is marked by a star. The upper
panel presents the local landscape in the Supergalactic plane,
SGX, SGY. Here objects with $\mid$SGZ$\mid >1.0$ Mpc were excluded. The bottom
panel shows the galaxy distribution in SGZ, SGY, where objects outside the
Sculptor filament with $\mid$SGX$\mid > 1.5$ Mpc are excluded.
The true, physical companions having the positive $\Theta_1$ are bound 
with their Main Disturber by solid lines, and some probable satellites 
with $\Theta_1 < 0$ are linked by dashed lines. Two dense clouds of
companions around the MW and M31 are delineated by ellipses.

The spatially nearest galaxy to KK 258 is a dwarf irregular galaxy 
UGCA 438 separated by 0.46 Mpc. This dwarf is a member of the NGC 55 
group consisting of 5 members: NGC 55, NGC 300, ESO 294-010, ESO 410-005,
and UGCA 438. Their radial velocities lie in the interval 
[ 53 -- 116 ] \kms. Two other neighbouring dwarf galaxies: IC 5152 and 
KK 258 with their radial velocities: 73 \kms and 150 \kms\ are distant associates
of the NGC 55 group. The Sdm galaxy NGC 55 has a
stellar mass of 3.0\,10$^9$ M$_\odot$. According to \citet{mk2011}, 
the NGC 55 group is characterized by a velocity dispersion of
only 26 \kms and the projected (virial) mass of 3.3\,10$^{11}$ M$_\odot$ (scaled
to the distance of 2.13 Mpc). Including in this group two its outliers
KK 258 and IC 5152 increases the velocity dispersion to 31 \kms
and increases the mean projected radius from 245 kpc to 347 kpc, yielding a virial
mass of 6.6\,10$^{11}$ M$_\odot$. Therefore, the virial mass-to-sum of stellar
mass of the NGC 55 group amounts to  6.6\,10$^{11}$/6.4\,10$^9$ = 103. 
As has been noted already by \citet{tully2006},
the association of dwarfs around the dwarf spiral NGC 55 has a
substantial amount of dark matter for the amount of light.

The properties of KK 258 as a representative of the dTr class 
are unlikely to have been caused by an interaction with a neighbour.
The status of the dwarf system KK 258 in the category 
of dTrs is difficult to ascribe to external, tidal 
effects. To explain the observed properties of the isolated transition 
type galaxy KK 258 plausibly one could invoke a
scenario of "cosmic web stripping" such as proposed by \citet{benitez2013}.
Perhaps though, in some fraction of cases, violent events in the early star forming phase drives most
gas out of the shallow potential of small galaxies and is never significantly replenished.

\section {Concluding remarks}

 In this paper we present a study of the KK~258 galaxy, an isolated transition
type dwarf system situated at a distance of 2.23$\pm$0.05 Mpc from the Milky
Way. The distance to KK~258 was measured by applying the tip of the red giant
branch method to V and I band images obtained with the Advanced Camera for
Surveys on the Hubble Space Telescope. We discuss the star formation history
of KK~258 derived from its colour-magnitude diagram. The main starburst in
KK~258 apparently occurred 12 -- 14 Gyr ago. The average rate of star formation
in that period was $7.9_{-3.2}^{+4.8}\times$10$^{-3}\,M_{\odot}/yr$. The
average metallicity of the stars formed is quite low, [Fe/H]$\simeq-1.6$.
About 70 per cent of the total stellar mass of the galaxy was formed during
the initial starburst. A weak but reliable sign of star formation is seen
also in the last billion years with an average rate of
$3.8_{-1.0}^{+3.0}\times$10$^{-4}\,M_{\odot}/yr$.

  Previous observations of KK~258 in the \ha{} line with the 6-meter Russian
telescope revealed a faint emissive patch of $\sim5$ arcsec size in the central
region of KK~258. The integral flux from the patch corresponds to the star
formation rate ${\rm log}[SFR] = -4.68$ in units of [M$_\odot$/yr]. This
emission knot is also seen in ultraviolet images obtained with the space
GALEX telescope yielding the star formation rate of ${\rm log}[SFR]=
-4.48$ M$_\odot$/yr in good agreement with the estimate from  \ha{} emission.

  Additionally, spectroscopic observations with the SALT telescope identify
KK~258 as possessing a heliocentric radial velocity $V_h = 92\pm5$ \kms{} or
the Local Group centroid velocity $V_{LG} = 150$ \kms{}. Under these velocity
and distance, KK~258 has a low peculiar velocity ($V_{pec} \sim 3$ \kms{})
with respect to the local Hubble flow with H$_0 = 73$ \kms{}\,Mpc$^{-1}$, 
perturbed by the Local Group with the radius of zero
velocity surface R$_0$ = 0.9 Mpc.

  We compared KK~258 with 29 other nearby dwarf galaxies classified in the
Updated Nearby Galaxy Catalog as been transition type. According to its
basic parameters: the absolute magnitude of $M_B = -10.3$ mag, the average
surface brightness of 26.0 mag\,arcsec$^{-2}$, the hydrogen mass of
${\rm log}(M_{HI}) < 5.75\,M_\odot$, and the specific star formation rate
of log[sSFR] = $-2.74$ KK~258  looks like a typical representative of the
sample of transition type dwarfs. However, unlike other dTr- galaxies,
KK~258 is sufficiently isolated that tidal interactions are not likely
to have affected its evolution in any way. From its spatial location KK~258
can be described as an outlier of the dwarf galaxy group around NGC 55,
which contains 5 members and resides in the scattered Sculptor filament.
The spatially nearest galaxy to KK~258 is a dwarf irregular galaxy UGCA~438
separated by 0.46 Mpc. The status of the dwarf system KK~258 in the category
of isolated transition dwarfs seems difficult to explain by usual gas
stripping mechanisms.

\section*{Acknowledgements}
This work is supported by RFBR grants 13-02-90407 and 13-02-92960.
LNM acknowledge the support from RFBR grant 13-02-00780 and Research 
Program OFN-17 of the Division of Physics, Russian Academy of Sciences.
Observations of KK~258 were made with Hubble Space Telescope in the course of program GO-12546.
Some of the observations reported in this paper were obtained with the 
Southern African Large Telescope (SALT),
DDT program \mbox{2013-1-RSA\_OTH-026}.
AYK acknowledge the support from the National Research Foundation (NRF) 
of South Africa.
We are grateful to Dmitry Makarov for help and discussion of the star formation
history of KK258.

\bibliographystyle{mn2e}
\bibliography{kk258}   

\bsp
\label{lastpage}

\end{document}

%% file: table1.tex
\begin{table*}
\caption{The Local Hubble flow galaxies}
\begin{tabular}{lccccrrrr}
\hline
Galaxy &     RA &     DEC &    D$_{MW}$ & Err & D$_{LG}$ & V$_h$  & V$_{LG}$ & $\Theta_1$ \\
       & \multicolumn{2}{c}{J2000.0} & \multicolumn{2}{c}{Mpc}  & Mpc     & km s$^{-1}$ & km s$^{-1}$ &   \\
  (1)  &           (2)    & (3)         &  \multicolumn{2}{c}{(4)}    &   (5)   &  (6)   & (7) & (8) \\
\hline
WLM= DDO221    & 00 01 58.1 &$-$15 27 40 &  0.97& 0.02 &  0.82 &$-$122 & $-$16& 0.0 \\
ESO349-031     & 00 08 13.3 &$-$34 34 42 &  3.21& 0.16 &  3.10 &  221 &  230  & 0.0 \\
NGC0055        & 00 15 08.5 &$-$39 13 13 &  2.13& 0.10 &  2.10 &  129 &  111  & 0.1 \\
NGC0300        & 00 54 53.5 &$-$37 40 57 &  2.15& 0.10 &  2.11 &  146 &  116  & 0.1 \\
NGC0404        & 01 09 26.9 &$+$35 43 03 &  3.05& 0.15 &  2.63 & $-$50 & 193  &$-$0.8 \\
HIZSS003       & 07 00 29.3 &$-$04 12 30 &  1.67& 0.17 &  1.79 &  288 &  108  &$-$1.1 \\
NGC2403        & 07 36 51.4 &$+$65 35 58 &  3.18& 0.16 &  3.15 &  125 &  262  & 0.2   \\
UGC04879       & 09 16 02.2 &$+$52 50 24 &  1.36& 0.06 &  1.34 & $-$25 &  33  &$-$0.7 \\
LeoA= DDO69    & 09 59 26.4 &$+$30 44 47 &  0.80& 0.04 &  0.96 &   24 & $-$40 &$-$0.1 \\
SexB= DDO70    & 10 00 00.1 &$+$05 19 56 &  1.43& 0.07 &  1.69 &  300 &  110  &$-$0.8 \\
NGC3109        & 10 03 07.2 &$-$26 09 36 &  1.32& 0.06 &  1.68 &  403 &  110  & 0.2  \\
SexA= DDO75    & 10 11 00.8 &$-$04 41 34 &  1.43& 0.07 &  1.74 &  324 &   94  &$-$0.8 \\
LeoP           & 10 21 45.1 &$+$18 05 17 &  2.00& 0.30 &  2.24 &  262 &  135  &$-$1.3 \\
NGC3741        & 11 36 06.4 &$+$45 17 07 &  3.22& 0.16 &  3.34 &  229 &  263  &$-$0.7 \\
DDO99          & 11 50 53.0 &$+$38 52 50 &  2.64& 0.14 &  2.74 &  251 &  257  &$-$0.8 \\
IC3104         & 12 18 46.1 &$-$79 43 34 &  2.27& 0.19 &  2.61 &  429 &  170  &$-$1.2 \\
DDO125         & 12 27 41.8 &$+$43 29 38 &  2.74& 0.14 &  2.81 &  206 &  251  &$-$0.8 \\
GR8= DDO155    & 12 58 40.4 &$+$14 13 03 &  2.13& 0.11 &  2.39 &  217 &  139  &$-$1.4 \\
UGC08508       & 13 30 44.4 &$+$54 54 36 &  2.69& 0.14 &  2.67 &   56 &  181  &$-$0.8 \\
DDO181= U8651  & 13 39 53.8 &$+$40 44 21 &  3.01& 0.15 &  3.08 &  214 &  284  &$-$1.1 \\
DDO183= U8760  & 13 50 51.1 &$+$38 01 16 &  3.22& 0.16 &  3.38 &  188 &  254  &$-$1.0 \\
KKH86          & 13 54 33.6 &$+$04 14 35 &  2.59& 0.13 &  2.89 &  287 &  209  &$-$1.4 \\
UGC08833       & 13 54 48.7 &$+$35 50 15 &  3.08& 0.15 &  3.24 &  221 &  280  &$-$1.1 \\
KK230= KKR03   & 14 07 10.7 &$+$35 03 37 &  2.14& 0.10 &  2.26 &   63 &  127  &$-$1.3 \\
DDO187= U9128  & 14 15 56.5 &$+$23 03 19 &  2.24& 0.12 &  2.43 &  160 &  180  &$-$1.4 \\
DDO190= U9240  & 14 24 43.5 &$+$44 31 33 &  2.80& 0.14 &  2.84 &  150 &  263  &$-$1.2 \\
KKR25          & 16 13 47.6 &$+$54 22 16 &  1.86& 0.12 &  1.79 & $-$79 &  128 &$-$1.0 \\
IC4662         & 17 47 06.3 &$-$64 38 25 &  2.44& 0.18 &  2.74 &  302 &  139  &$-$1.3 \\
SagdIr=ESO594-4& 19 29 59.0 &$-$17 40 41 &  1.04& 0.07 &  1.14 & $-$79 &   21 &$-$0.4 \\
DDO210=Aquarius& 20 46 51.8 &$-$12 50 53 &  0.98& 0.05 &  0.97 &$-$140 &   11 &$-$0.3 \\
IC5152         & 22 02 41.9 &$-$51 17 43 &  1.97& 0.12 &  2.08 &  122 &   73  &$-$1.3 \\
KK258=ESO468-20& 22 40 43.9 &$-$30 47 59 &  2.23& 0.11 &  2.34 &   92 &  150  &$-$1.1 \\
Tucana         & 22 41 49.0 &$-$64 25 12 &  0.88& 0.04 &  1.09 &  194 &   73  &$-$0.2 \\
UGCA438        & 23 26 27.5 &$-$32 23 26 &  2.18& 0.12 &  2.11 &   62 &   99  &$-$0.4 \\
KKH98          & 23 45 34.0 &$+$38 43 04 &  2.52& 0.13 &  2.11 &$-$132 &  156 &$-$0.9 \\
\hline
\end{tabular}
\end{table*}

%% file: table2.tex
\begin{table*}
\caption{List of 30 transition type dwarf galaxies within 5.0 Mpc}
\begin{tabular}{lcccrrrrrrrr}
\hline
 Name     &        J2000.0 &    D & meth & M$_B$ & SB & lg(M$_*$)& $lg(M_{HI}$ & lg(SFR) & lg(sSFR)& $\Theta_1$ &Group \\
   & &   & &  &  & & $/M_*)$ &  & &  & \\
 (1)      &          (2)   &   (3)& (4)  & (5)   & (6)&    (7)      &    (8)          &  (9)    & (10)     &  (11)    & (12) \\
\hline
ESO410-005   & 001531.4$-$321048 & 1.92 & TRGB & $-$11.6 & 24.5 & 6.88  &  $-$0.97 &  $-$3.97 &  $-$10.85 &    0.0 &  N55 \\
Cetus        & 002611.0$-$110240 & 0.78 & TRGB & $-$10.2 & 26.3 & 7.02  & $<-$2.84 & $<-$6.54 & $<-$13.56 &    0.3 &  M31 \\
ESO294-010   & 002633.3$-$415120 & 1.92 & TRGB & $-$10.9 & 25.0 & 6.25  &  $-$0.77 &  $-$3.86 &  $-$10.11 &    0.4 &  N55 \\
KDG002       & 004921.1$-$180428 & 3.40 & TRGB & $-$11.4 & 25.4 & 6.81  &  $-$0.86 &  $-$3.87 &  $-$10.68 &    0.5 &  N253 \\
LGS 3        & 010355.0+215306   & 0.65 & TRGB & $-$9.3  & 25.4 & 5.96  &  $-$0.94 &  $-$5.23 &  $-$11.19 &    1.5 &  M31 \\
Phoenix      & 015106.3$-$442641 & 0.44 & TRGB & $-$9.6  & 26.0 & 6.08  & $<-$0.86 &  $-$4.43 &  $-$10.51 &    0.7 &  MW \\
UGC01703     & 021255.8+324851   & 4.19 & SBF  & $-$11.5 & 24.9 & 7.56  & $<-$1.26 & $<-$4.85 & $<-$12.41 & $-$1.3 &  -- \\
$[$KK2000$]$ 03  & 022442.7$-$733046 & 4.10 & mem  & $-$12.3 & 26.4 & 8.21  & $<-$1.93 & $<-$5.02 & $<-$13.23 & $-$0.8 &  -- \\
KKH22        & 034456.6+720352   & 3.50 & mem  & $-$11.4 & 25.8 & 6.80  & $<-$0.34 & $<-$4.05 & $<-$10.85 &    1.1 &  IC342 \\
CKT09d0926+70& 092627.9+703024   & 3.40 & TRGB & $-$9.6  & 25.1 & 6.11  & $<-$0.59 &  $-$4.30 &  $-$10.41 &    1.7 & M81 \\
CKT09d0939+71& 093915.9+711842   & 3.75 & TRGB & $-$8.4  & 25.8 & 5.63    &    --    &  $-$5.76 &  $-$11.39 &    2.0 &  M81 \\
CKT09d0944+71& 094434.4+712857   & 3.36 & TRGB & $-$11.1 & 25.0 & 7.38  & $<-$1.90 &  $-$5.74 &  $-$13.12 &    1.6 &  M81 \\
Antlia       & 100404.0$-$271955 & 1.32 & TRGB & $-$9.8  & 26.1 & 6.49  &  $-$0.55 &  $-$3.56 &  $-$10.05 &    2.3 &  N3109 \\
KDG063       & 100507.3+663318   & 3.50 & TRGB & $-$12.1 & 25.9 & 7.80  &  $-$0.89 & $<-$5.02 & $<-$12.82 &    2.0 &  M81 \\
KDG064       & 100701.9+674939   & 3.70 & TRGB & $-$12.6 & 25.4 & 7.98    &    --    &  $-$5.52 &  $-$13.50 &    2.7 & M81 \\
CKT09d1014+68& 101455.8+684527   & 3.84 & TRGB & $-$7.9  & 25.2 & 6.12  & $<-$0.43 &  $-$4.73 &  $-$10.85 &    2.0 & M81 \\
CKT09d1015+69& 101506.9+690215   & 3.87 & TRGB & $-$7.6  & 26.0 & 6.01    &    --    &  $-$5.43 &  $-$11.44 &    1.3 & M81 \\
HS117        & 102125.2+710658   & 3.96 & TRGB & $-$11.2 & 26.5 & 6.72  &  $-$1.71 &  $-$4.32 &  $-$11.04 &    1.2 &  M81 \\
KDG090       & 121457.9+361308   & 2.86 & TRGB & $-$11.5 & 25.4 & 6.86  & $<-$1.30 & $<-$5.44 & $<-$12.30 &    1.3 &  N4214 \\
$[$KK2000$]$ 51  & 124421.5$-$425623 & 3.60 & mem  & $-$10.9 & 24.4 & 6.60  & $<-$0.28 & $<-$5.01 & $<-$11.61 &    0.8 & N5128 \\
KK166        & 124913.3+353645   & 4.74 & TRGB & $-$10.8 & 25.1 & 7.28    &    --    & $<-$5.02 & $<-$12.30 &    0.7 &  N4736 \\
KDG218       & 130544.0$-$074520 & 5.00 & txt  & $-$11.9 & 26.6 & 7.71  & $<-$1.10 &  $-$4.85 &  $-$12.56 & $-$1.0 &   -- \\
NGC5011C     & 131311.9$-$431556 & 3.60 & mem  & $-$14.1 & 22.9 & 7.95  & $<-$0.83 &  $-$3.39 &  $-$11.34 &    2.0 &  N5128 \\
$[$KK2000$]$ 54  & 132132.4$-$315311 & 4.60 & mem  & $-$10.5 & 25.7 & 6.44  & $<-$0.66 & $<-$4.87 & $<-$11.31 &    0.9 &  N5236 \\
KK198        & 132256.1$-$333422 & 4.60 & mem  & $-$11.0 & 24.7 & 7.34  & $<-$1.52 &  $-$4.26 &  $-$11.60 &    0.8 &  N5236 \\
AM1320-230   & 132329.9$-$232335 & 4.60 & mem  & $-$11.1 & 25.2 & 7.41  & $<-$1.63 &  $-$4.32 &  $-$11.73 &    0.4 &  N5236 \\
KK203        & 132728.1$-$452109 & 3.60 & mem  & $-$10.5 & 24.6 & 6.46  & $<-$0.14 & $<-$4.97 & $<-$11.43 &    2.0 &  N5128 \\
And XXVIII   & 223241.2+311258   & 0.65 & TRGB & $-$7.7  & 26.3 & 6.04     &    --    & $<-$6.60 & $<-$12.64 &    1.1 &  M31 \\
KK258        & 224043.9$-$304759 & 2.23 & TRGB & $-$10.3 & 26.0 & 7.16  & $<-$1.41 &  $-$4.48 &  $-$11.64 & $-$1.1 &   -- \\
Tucana       & 224149.0$-$642512 & 0.88 & TRGB & $-$9.2  & 26.5 & 6.62  & $<-$2.44 &  $-$6.04 &  $-$12.66 & $-$0.2 &  MW \\
\hline
 Median      &         --        & 3.5  &  --  &  --10.9 & 25.4 &  6.80  & $<-$1.0  &   $<-$4.8 &  $<-$11.7  &    0.9 &   -- \\
\hline
\end{tabular}
\end{table*}